\documentclass[journal,twocolumn]{IEEEtran}

\usepackage{url}
\usepackage{graphicx}

\usepackage{amsfonts}
\usepackage{multicol}
\usepackage{booktabs}
\usepackage{textcomp}
\usepackage{comment}

\usepackage{threeparttable}
\usepackage{setspace}

\usepackage{stfloats}
\usepackage[below]{placeins}

\usepackage{cite}

\usepackage{pifont}

\usepackage{amssymb}

\usepackage{multirow}
\usepackage{amsmath}
\usepackage{csquotes}
\usepackage{color,soul}

\usepackage{comment}

\ifCLASSINFOpdf
\else
\fi

\begin{document}

\title{Challenges of AI in Wireless Networks for IoT}

\author{\IEEEauthorblockN{Ijaz Ahmad\IEEEauthorrefmark{1},
Shahriar Shahabuddin\IEEEauthorrefmark{2},
Tanesh Kumar\IEEEauthorrefmark{3},
Erkki Harjula\IEEEauthorrefmark{3},
Marcus Meisel\IEEEauthorrefmark{4},
Markku Juntti\IEEEauthorrefmark{3},
Thilo Sauter\IEEEauthorrefmark{4}\IEEEauthorrefmark{5},
Mika Ylianttila\IEEEauthorrefmark{3}
\\}
\IEEEauthorblockA{\IEEEauthorrefmark{1}VTT Technical Research Center of Finland, Espoo, Finland }
\IEEEauthorblockA{\IEEEauthorrefmark{2}Nokia, Mobile Networks, Oulu, Finland }
\IEEEauthorblockA{\IEEEauthorrefmark{3}Centre for Wireless Communications, University of Oulu, Finland }
\IEEEauthorblockA{\IEEEauthorrefmark{4}Institute of Computer Technology, TU Wien, Austria}
\IEEEauthorblockA{\IEEEauthorrefmark{5}Center for Integrated Sensor Systems, Danube University Krems, Austria}
}

\maketitle


\begin{abstract}

The Internet of Things (IoT), hailed as the enabler of the next industrial revolution, will require ubiquitous connectivity, context-aware and dynamic service mobility, and extreme security through the wireless network infrastructure. Artificial Intelligence (AI), thus, will play a major role in the underlying network infrastructure. However, a number of challenges will surface while using the concepts, tools and algorithms of AI in wireless networks used by IoT. In this article, the main challenges in using AI in the wireless network infrastructure that facilitate end-to-end IoT communication are highlighted with potential generalized solution and future research directions.

\end{abstract}

\IEEEpeerreviewmaketitle


\section{Introduction}

Internet of Things (IoT), the term first coined by Kevin Ashton in~\cite{IOT}, is an extension of network connectivity to physical devices, such as actuators, sensors and mobile devices, enabled to interact and communicate among themselves, and can be controlled or monitored remotely. IoT, hailed as the enabler of the next industrial revolution, will transform how we view, interact and use the current physical systems available around us. It already have major impacts on health care, smart-homes, manufacturing, commerce, education and many other key areas of the daily life. The IoT market is undergoing incredible growth and the IoT industry is projected to grow tenfold by 2025~\cite{IoTAnalytics}. With smart cities in a foreseeable sight having automated IoT in various forms, such as Unmanned Aerial Vehicles (UAVs), smart-homes, e-health devices, and context-aware Augmented Reality (AR) and Virtual Reality (VR) applications used in daily routines, the underlying communication networks must evolve to meet their needs. Communication networks must also support autonomous operations due to the continuously changing services, unprecedented increase in network traffic, and increasingly complex security threat landscape due to the amalgamation of diverse IoT devices and services. All these challenges further add into increasing the complexity of network operations.

Artificial Intelligence (AI) with its disciplines, i.e., machine learning (ML), is the primary enabler of an autonomous and intelligently operating network. Since the groundbreaking work of Hinton {\it et al.}~\cite{hinton2006fast} in 2006 on a fast training method for deep neural networks, there has been a reinvigorated interest on neural networks and other ML methods in communication networks~\cite{8360430}. The application of ML in wireless networks has been of immense interest and a plethora of research articles has been published. However, this is just not the first age of AI where it has attracted a huge attention of research community. During 70s and 80s, there have been immense enthusiasm and optimism on AI in cycles, which was followed by periods of AI winters, a term coined to explain low interest in AI. The current era of AI is bolstered by advanced semiconductor technologies and the advent of cloud and distributed computing. In spite of all these technological advancements, a number of challenges still remain today in order to successfully deploy AI based solutions on a competitive basis in wireless networks. Instead of considering AI as an omnipotent solution, a cautious approach and a careful comparison against state-of-the-art solutions is necessary to make the AI-based solutions applicable and successful in future communication networks.

For capitalizing on IoT, having increasing number of connected diverse devices with emerging smart services, autonomous network operations leveraging AI is inevitable. For example, the conglomeration of heterogeneous IoT devices in UAVs, e-health, manufacturing, AR/VR, wearables, and smart homes through the communication technologies will make it very difficult to differentiate a security attack from legitimate traffic, and may not be practically possible or manageable without using AI~\cite{moh2018machine}. Therefore, autonomous network operations are contemplated to be possible with embedding and using the concepts, technologies and algorithms of AI in wireless networks. To avoid repeating the definitions of the vast number of types and disciplines of AI, in this article the term AI is referred to techniques that are used to i) gather (raw) data from the network environment, ii) perform computation on it (e.g. for classification, training and testing), and iii) produce intelligent actionable information for the network. This may include the required systems of supervised, unsupervised or semi-supervised learning, to name a few.

However, using AI in wireless networks will bring its own challenges, which may not be worth considering in other fields such as machine vision and robotics, but highly important in communication networks, specifically in the case of IoT~\cite{7738452}. For example, gathering the raw data for training the system incurs network overhead. Storing the raw data requires storage systems, and in big data, big storage systems are required. Similarly, performing computation on the data to extract actionable information requires higher computing resources. If resources are available in high-end servers in centralized cloud systems, latency critical applications will be challenged by the communication latency, besides other factors~\cite{8865093}. In decentralized systems, sharing data and training models or parameters of AI algorithms will not only require higher communication network resources, but also open security challenges. Hence, using AI in wireless networks has many challenges that are not counted in most of the research in this direction.  

Most of the state-of-the-art research articles attempt to solve specific challenges using AI in wireless networks while ignoring the resulting challenges arising as a consequence. Therefore, major challenges that arise due to using AI in wireless networks are discussed in this article, mainly from the point of view of IoT. The main purpose of highlighting the challenges is twofold. First, to grasp research attention to the limitations of AI from the perspectives of wireless networks. For example, wireless channels are prone to errors, data distribution can be non-uniform keeping in mind the possibility of unavailability of data due to various reasons such as jamming attacks, and wireless networks can have limited capacity such as bandwidth, storage and computing required for AI. Second, to motivate further research on developing AI-based solutions that are either specific to wireless networks or avoid facing situations where solving one problem creates another in the wireless network infrastructure. For example, learning from the big data generated by IoT with the help of AI in the edge might yield the required results, however, the required storage and processing might be too costly compared to its benefits. Therefore, how to avoid pitfalls in using AI in future wireless networks, specifically in the case of IoT, is the main theme of this article.

\section{Challenges Posed by AI in the Wireless Network Infrastructure}

\begin{figure*}[h]
\centering
\includegraphics[keepaspectratio,width=1.60\columnwidth]{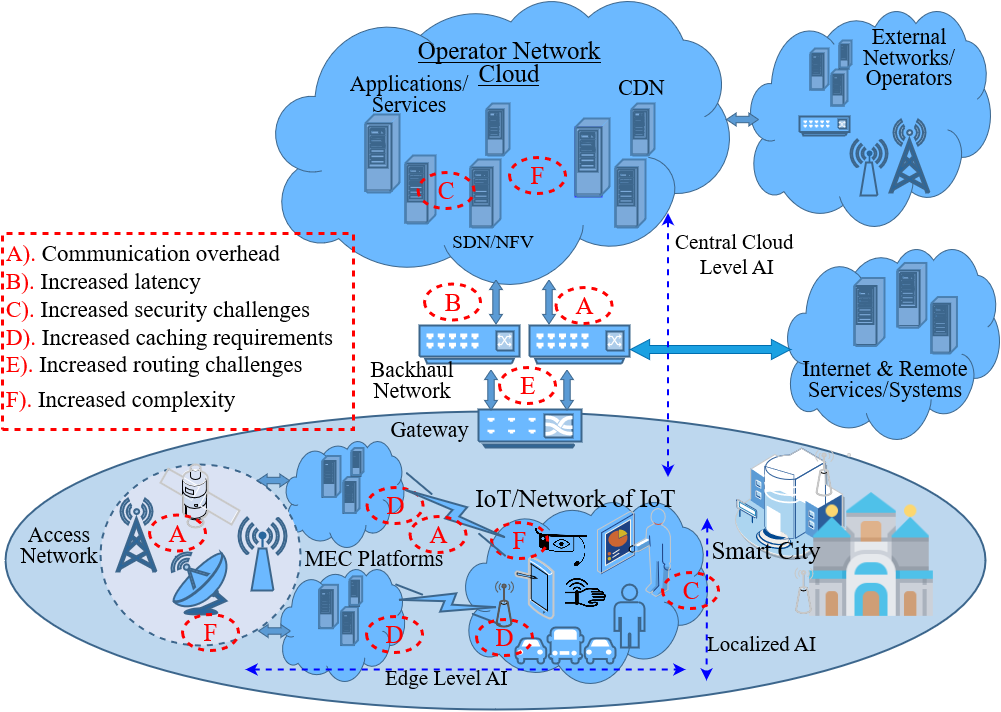}
\caption{A generic network architecture showing few key areas of AI in IoT and corresponding challenges.}
\label{Fig11}
\end{figure*}

To complement for resource limitations, heterogeneity, and complexity in IoT on one hand, and big data on the other hand, various concepts of enhanced computing, storage, link, and bandwidth are bundled with the concepts, tools and algorithms of AI. Therefore, huge research efforts are going on in this direction as presented in~\cite{8373692, 8110603}. Moreover, new concepts and disciplines of AI in different network systems or network services are proposed, discussed, and evaluated continuously~\cite{7792374}. Fig.~\ref{Fig11} presents a generic global network in which AI is used in different segments, including IoT devices, and the network that connects diverse IoT devices. However, a number of challenges will surface in the when AI is used but proper consideration is not given to the underlying network architecture and infrastructure. In this section, we discuss the main challenges that will be on the forefront when AI is used in future wireless networks. The most common challenges, related to almost all types of networks as depicted in red in Fig.\ref{Fig11}, are described below.  

\subsection{Higher communication overhead}

\begin{figure}[h]
\centering
\includegraphics[keepaspectratio,width=0.55\columnwidth]{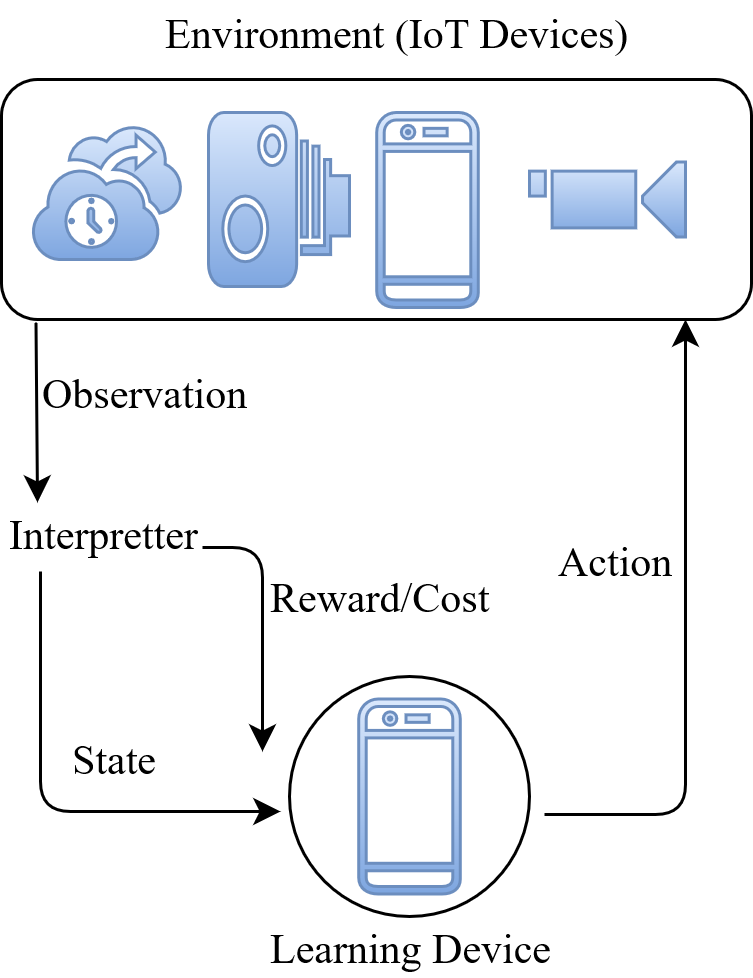}
\caption{A simplified reinforcement learning system.}
\label{FigRL}
\end{figure}

Using AI to improve the efficiency of IoT devices, services offered by IoT devices, or to improve the functionality of the underlying network used by IoT will have additional communication overhead. The communication overhead caused by AI can be attributed to the very basic operating principles of AI systems. For example, ML systems derive useful information from (large scale) data that needs to be communicated between the devices running ML algorithms. To illustrate the extra communication costs, consider the learning device in Fig.~\ref{FigRL} using communication bandwidth and spectrum for observation, communicating results of the interpreter, and then sharing the action space with other IoT devices in the environment. Thus, the communication costs of learning algorithms in ML can be generically determined by i) number of communication rounds required to observe or learn the environment (ML algorithm convergence), ii) number of channels used per communication round, and iii) bandwidth or spectrum used per channel in a communication round. 

In IoT, the huge amount of diverse data generated by the massive number of connected IoT devices will require very high memory and processing resources~\cite{7906512}. Thus, the limited capacity of IoT devices will force the ML processing and required storage to other available resources, the most prominent being the edge nodes, mainly being near to IoT devices to meet the latency requirements~\cite{8270639}. However, it is highly challenging in most use-cases of big data generated by IoT to fit and process the entire data set in the edge as discussed in~\cite{8270639}. Therefore, two approaches will most likely be used, first, coordinated distributed processing in multiple edge nodes, and second, centralized processing in a pool of larger processing and storage systems such as centralized cloud systems. In both cases, the communication overhead will increase much higher than anticipated. 

For distributed coordinated processing of ML in multiple edge nodes, distributed ML, called federated learning, is proposed~\cite{konevcny2016federated}. In federated learning, the training data remains distributed over a large number of nodes. A centralized model is trained by the distributed nodes performing computation over their individual data independently~\cite{konevcny2016federated}. Most distributed ML systems usually contain a group of server nodes that manage global parameters, and worker nodes that pull the latest parameters and push the gradients to server nodes for update operation. This process of pulling and pushing the parameters and gradients during the training cause huge network traffic as demonstrated in~\cite{7823861}.

Different ML algorithms for federated learning are evaluated in~\cite{8664630} that reveal astonishing results. For example, when the data distribution is non-uniform, the convergence time of the training model is very high and the accuracy is very low. Therefore, synchronization of a distributed ML system is critical in order to accurately update each system to the global model using fresh information. However, synchronization is a high-cost operation that require significant communication rounds for fresh updates for all the participating nodes. Heterogeneity in resource capacity of IoT devices, and diversity in data sets will further increase the communication rounds in order to synchronize ML among all the participating nodes~\cite{8664630, Arjevani2015}. In IoT, as the data sets for learning grow larger, the models will be more complex and training AI models will increasingly require distributing the model optimization parameters over multiple machines~\cite{konevcny2016federated}. As a result, using AI in large IoT networks with multiple diverse nodes and heterogeneous links will result in very high communication overhead costs~\cite{8489985}.

\textbf{\textit{The Case of Centralized Cloud Systems:}} Computation and storage costs can be minimized by using centralized cloud-based systems for ML. However, the communication costs in bringing raw data, training model parameters, and later the outcome of the learning algorithms between the cloud and end-user devices will consume high link and bandwidth budgets. Traditional AI algorithms are designed for highly controlled environments such as data centers, where the assumptions are that i) the data is independent and identically distributed (i.i.d) among machines, and ii) high throughput networks are always available. In wireless networks, both of these assumptions may not always hold true, requiring frequent re-sending of data leading to further network resource dedication. Similarly, end-to-end security procedures might further increase the network overhead. In summary, for AI-based operations, continuous gathering of data will be inevitable, dissemination of decisions must happen, and scaling network resources (e.g. channels and bandwidth, and access and backhaul networks, etc.) will pose significant challenges, and more so in the case of AI in centralized cloud systems.   

\subsection{Challenges for Latency-Critical IoT Systems}

The dynamic nature of future IoT services will require real-time computation, ideally near the users, or otherwise with no observable delays~\cite{Ahmad2018}. However, due to low capacity IoT devices will take considerably longer time for AI processing within IoT devices. In the case of processing in the edge, it is concluded by Arjevani {\it et al.}~\cite{Arjevani2015}, that many communication rounds will be required and still provide worst-case optimum in minimum assumption situations. In simple words, (raw) data acquisition, then data analysis and training, and the continuous feedback loop in ML will introduce much higher delay. Even traditional (non-ML) iterative or feedback systems are having challenges in terms of computation and link delays to meet the real-time requirements of dynamic services and highly mobile users~\cite{8227766}. Whereas, the delay in training ML models, for instance in streaming applications, will make it very difficult to match the latency requirements~\cite{8279592}.

An interesting scenario of V2X using federated learning is evaluated in~\cite{8647927}. The results reveal that in most cases, even near the user scenarios, the federated learning approach incurs higher latency. In latency-critical systems such as V2X communication and tele-surgery, an extremely small delay, for instance in moving the steering wheel or robotic arm, can be catastrophic. In~\cite{OffloadIoT}, a CNN-based object-inference task was offloaded to cloud which leads to a 2\,s to 5\,s latency. The experiments were run in US and China and the authors concluded that the variability of latency makes the service unreliable. Therefore, continuous learning and adjustment of systems using the apparatus of AI for such critical operations must ensure latency first. 

Furthermore, the limitations in time for usefulness of data for AI processing, and validity of the outcome of AI mechanisms must be counted. To elaborate these limitations, consider intrusion detection systems. Analysis of data for intrusion detection is highly time-sensitive, and if the communication medium introduces delay, the whole process might be rendered useless~\cite{Suthaharan}. Investigating distributed ML systems,~\cite{7823861} reveals that network communication consumes more time by an order of magnitude than computation to train ML models. Similarly, the completeness of data required within the time frame to make observations on is also crucial. The question holds true for observation-oriented or data-oriented decision-making systems of any kind. Therefore, latency in such systems is as crucial as the validity or accuracy of the system. This can be further clarified with the example of object recognition through deep learning, as described in~\cite{8057306}. Images for object recognition tasks processed locally consume seven W energy, and when processed in cloud it consumed two W energy. However, the latency goes well beyond the constraints of 500 ms. It took between two seconds to five seconds of time when processed in the cloud. Therefore, it is concluded in~\cite{8057306} that for real-time deep learning tasks, cloud is not yet a viable solution due to higher latency.

\textbf{\textit{The Case of Industrial Control Systems:}} Industrial Control Systems (ICSs) have very complex requirements such as low latency, high reliability, security and safety, and 5G seems to be promising from many aspects as highlighted in~\cite{7883994}. The requirements of ICSs are different than other systems and services, for example, the bandwidth requirements for data transmission can be as low as few bytes, whereas the latency requirements for real-time control messages in production and manufacturing can be as strict as 250 micro seconds~\cite{7842415}. Albeit ICSs are moving towards distributed automated systems, distributed systems connected through communication networks have a considerable delay mainly due to the inter-working architectures from access to core networks. Only the core network in 4G cellular system introduces a 39 ms delay to contact the gateway towards the Internet~\cite{7842415}. The measurements reported in~\cite{7842415} are in two cells and low traffic scenario. In peak hours, the delay can further grow up to 85ms. Even though the delay can be minimized for instance by localizing various network functions in the Edge in 5G, no significant change is made for traffic reaching outer networks. Hence, adding ML in ICSs will add further delay which raises serious concerns about the benefits of ML in ICSs.

\subsection{Challenges in Routing and Network Traffic Control}

Even though AI has been proposed for routing, traditional AI/ML techniques such as artificial neural networks have evident shortcomings in terms of scalability and computation efficiency when considered for routing~\cite{7792369}. Measuring the benefits of using deep learning-based routing vs traditional OSPF routing mechanism in~\cite{7935536}, the results reveal that OSPF yields the same throughput and average delay when the signaling interval between routers is more than a certain threshold. However, counting the computational and storage resources, straightforward OSPF is a better option for the core and backhaul networks, where changes are less likely compared to the dynamics in access networks. Furthermore, mutating or changing IP addresses or packet header fields for either security attacks, or preventing security attacks~\cite{7185417} will further challenge the phenomenon of learning, and may lead to continuous feedback loops for finding the best route.

A lot of research efforts are dedicated to using AI in dynamic networks. Dynamic networks have frequently changing topologies that require frequent sharing of information among nodes in the network. An example of dynamic networks is MANETs, which are composed of resource constrained mobile devices. MANETs are formed randomly and spuriously by freely moving nodes. Thus, the routing protocols usually have higher overhead due to dissemination of topology information, as well as sharing information because of transient disruptions during routing protocol convergence~\cite{Suchara:2011:NAJ:1993744.1993756}. However, the constantly changing topologies lead to continuous arrival of new information. Such systems behave like a closed loop system making it hard for the learning algorithms to converge within the latency constraints. 

\textbf{\textit{The Case of Software Defined Networking:}} Since traditional network traffic control systems heavily rely on pre-defined policies hardwired in the data plane devices, new solutions such as Software Defined Networking (SDN) have been sought to minimize manual configurations and enable run-time changes in network policies. SDN splits the network control-data planes, centralizes the network control plane, and enables programmability of the network equipment. Thus, SDN enables dynamicity in communication networks, which is required in wireless networks to cope with sudden changes in user behavior, network traffic, and air interfaces. Therefore, ML-based management of complex network systems, and ML-based route selection in SDN, according to the traffic requirements of different applications have been proposed in~\cite{7217798}, and~\cite{8077095} respectively. Hence, AI-based network traffic control in SDN has gained research attraction recently mainly to cope with the dynamicity of mobile nodes, diverse services and increasing traffic variations.

Even though SDN provides promising solutions to many challenges, it has its own inherent challenges of scalability and security, mainly due to the centralized control architecture~\cite{7226783}. In simple words, the centralized SDN controllers need to be scalable enough to install flow rules in the entire data plane under its control within latency constraints. In terms of resilience,~\cite{6076935} reveals that it is hard to achieve carrier grade requirement of restoration within 50ms in large OpenFlow networks. However, using AI in SDN will require either adding software modules to the controller or adding an application on top of the control plane. In both situations, the controller involvement in the data plane will further increase by consistently feeding information (e.g., flow patterns, flow statistics, or samples of packets) to AI algorithms. Hence, using AI in SDN without giving proper consideration to its inherent limitations will further increase its challenges.

\subsection{Challenges in caching}

Network caching systems temporarily store data or information near the users in order to minimize redundant network traffic~\cite{6736753}. Traditionally a router, for example, would cache data that has higher requests or frequently passes through it. However, the explosion of big data from IoT will really challenge the fundamentals of in-network caching. AI-based system have been proposed to enable the network to learn which data or information to cache~\cite{8403948}. However, using AI within the network devices, e.g. routers and switches, will consume resources meant for storing routing procedures and paths, and access control lists, etc. For example, in~\cite{8172025} the authors proposed content caching using deep learning in SDN. Considering the OpenFlow standard of SDN used in the analysis, OpenFlow switches have limited capacity to store unsolicited flows until the controller updates the flow tables, and in some cases have limited capacity to store flow rules~\cite{7226783}. Furthermore, the SDN controllers have serious scalability challenges, and therefore various hierarchical and distributed control plane architectures have been proposed, as described in~\cite{7226783}. Albeit these limitations, the authors in~\cite{8172025} suggest sending the prediction output of the deep learning algorithm to the controller so that the controller knows popularity of the contents in the network it manages. The humongous increase in the number, types and services of IoT will increase the amounts and types of popular content. Hence, using AI algorithms on the content within the network will require a drastic increase in memory size, as well as processing capability to meet the requirements of real-time services. Therefore, content caching in the edge is proposed that has its own limitations and challenges as described below.

\textbf{\textit{The Case of Edge Resources vs Data Growth:}} Partial or full storage, and processing in the edge is proposed to deal with varying and massive amount of data under the constraints of time-validity or duration, e.g., for useful information retrieval from raw data, and generating actionable information or intelligence. However, the main question, usually ignored, is that how much storage and processing will be required? Many evaluations of edge-enabled deep learning, such as discussed in~\cite{8270639}, consider the maximum data size of a task as low as 1 Mbps and increases the number of edge nodes for processing the data by many numbers at a time (10-90 for 1000 tasks). Having said that, the user experienced data rates in 5G are expected to be 1 Gbps in downlink, 500 Mbps in uplink, and capacity targets can be as high as 15Tbps/km2 with 250 thousand user devices in a square kilometer~\cite{3GPPdatarates}. Currently, the data size of medium-level operators easily exceeds 100s of terabytes, and will further increase since video traffic (4K, 8K, 3D video, 360-degree video) will account for around 75\% of traffic by 2023 according to the GSM alliance. For example, the AT\&T network carries more than 200 petabytes a day. Keeping these facts in mind, the main challenge in the edge is the computation needed for real-time analysis of raw data generated by end-user devices and IoT, mainly due to the diversity of applications generating different traffic. Traditional ML, however, requires full access to data sets with centralized computing through ultra-fast chipset, Graphics Processing Units (GPUs), connected through up to 256 Gbps connections. Thus, specific processing units, such as tensor processing units~\cite{8192463} are required that will be capable of matching with the quantity of data passing through the networks. Keeping such huge amounts of data within the networked devices, or even in edge nodes for AI processing will be highly challenging.

\subsection{Security and Privacy Challenges}

The application of AI for IoT security has got a lot of momentum in recent years. AI is typically used for discovering a pattern in existing data, detecting outliers, predicting values or feature extraction which are all very crucial tools to secure IoT devices and network. The main objective of using AI for IoT security is detecting a security breach which can be divided into three categories according to~\cite{moh2018machine}: (1) malware detection, (2) intrusion detection, (3) data anomaly detection. For example, in~\cite{ham2014linear}, the authors presented a linear SVM based android malware detection for reliable IoT devices. An example of intrusion detection can be found in~\cite{NbayesIoT} where the authors applied a two tier classification mechanism based on Naive Bayes and \textit{K}-Nearest Neighbor to prevent intrusion detection of an IoT network. An example of data anomaly detection is presented in~\cite{canedo2016using} where the authors propose using ANN in an IoT gateway to detect anomalies in the data sent from the edge devices. We invite interested readers to go through~\cite{moh2018machine} to learn more about the AI schemes for security purposes.

A key question for using AI in the context of IoT security is how to generate a high-quality training dataset containing possible attack types and patterns. A high quality training dataset is essential for the accuracy of AI schemes. A diverse training dataset containing information that reflects all the strategies of real world attacks is required for successful deployment of AI methods for IoT security. However, due to a large number of devices generating large volumes of data, a real-time high quality data streaming and extraction remains a challenge. In addition, extracting a reliable dataset through collaboration of different devices can also be challenging due to a wide diversity of IoT devices. 
Most publications on AI for IoT security are applied for high-quality data. For example, the intrusion detection mechanism 
of~\cite{NbayesIoT} use NSL-KDD dataset to train and validate the AI scheme. However, the NSL-KDD may not be a perfect representative of existing real networks. 
Due to the amalgamation of a large number of heterogeneous devices in an IoT network, the effect of noise and interference can corrupt a dataset. Therefore, AI methods based on high-quality datasets to secure IoT are highly infeasible. It should be noted that  
acquiring dataset for training in the context of IoT security is more difficult than for image or natural language processing.

ML techniques such as supervised learning, unsupervised learning, and reinforcement learning based approaches for IoT authentication, access control, secure offloading, and malware detection schemes are studied in~\cite{8454402}. The authors conclude that both supervised and unsupervised learning methods for IoT security have serious challenges of oversampling, lack of sufficient training data, and bad feature extractions. Supervised learning-based intrusion detection systems have, sometimes, miss-detection rates that cannot be neglected in IoT systems. RL-based system can cause network disaster for IoT systems at the beginning stage of learning, i.e., exploring bad security policies to achieve optimal strategies. The optimal solution in such cases is to have backup security mechanisms to protect IoT systems during the exploration stage of the learning processes~\cite{8454402}.

Another key challenge is the inherent security flaws of traditional AI mechanisms. Firstly, adversaries can feed polluted training data during training and reduce the performance of AI schemes. This attack is commonly known as poisoning attack. Secondly, an adversary can feed feasible new inputs in an attempt to evade detection, which is known as evasion attack. Thirdly, adversaries can create their own AI models by public API and refine their own model using it as a guide. Therefore, security of an AI scheme itself needs to be taken into account before using it to secure IoT systems and devices. Several techniques such as, data sanitization, adversaries retraining and homomorphic encryption exists to make an AI scheme more secure against its inherent security flaws. The security flaws of AI algorithms and their countermeasures are presented in detail in~\cite{guan2018machine}.

In~\cite{5504793}, some basic questions are put forward regarding the use of AI in security systems. Comparing the use of AI in other disciplines, Sommer {\it et al.}~\cite{5504793} state that it is not only harder to use AI for intrusion detection, but the premise of using AI to find novel attacks does not hold true. The reason is simple; AI algorithms typically use previous experiences or knowledge to build decisions upon, whereas, for novel attacks the system may not have prior data or information available. Another key question is how to intervene once an IoT device is discovered to be part of a DDoS attack. Removing the device from the network might not be possible, mainly if it is a critical device. Most AI methods just focus on detecting an attack and do not address the mechanism of rectifying this situation.

\textbf{\textit{The Case of Privacy:}} Privacy with its all potential colors such as legal, ethical, moral, is likely to get exposed in the era of AI controlled networks. For example, in~\cite{6871674} the authors propose to harness user behavior, social relationships, and other personal attributes from social networks for proactive caching in the edge. Similarly, AI algorithms themselves can leak sensitive information when they are subjected to security attacks~\cite{10.1007/978-3-319-99740-7_1}. The adversary can perform inverse operation to attain the input data, such as, patient medical information, user fingerprint or customer purchase record. Therefore, preserving privacy in the age of AI will be challenging from both the algorithmic security and human invasions perspective. There are various approaches now emerging to safeguard privacy of user data, such as differential privacy models~\cite{guan2018machine}, and encoding and shuffling algorithms~\cite{Bittau:2017:PSP:3132747.3132769}. However, privacy requires more regulatory efforts as well, since in most cases the privacy challenges arise on the operator or service provider sides~\cite{8712553}.

\subsection{System Complexity Challenges}

\begin{figure*}[h]
\centering
\includegraphics[keepaspectratio,width=2.0\columnwidth]{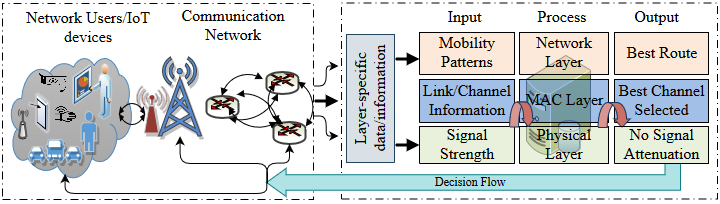}
\caption{An example of layer-wise AI implementation.}
\label{Fig3}
\end{figure*}

Deploying AI in communication networks will further increase the complexity of the system, if the implementation is carried out as we see currently: implement case-specific ML to achieve one objective, ignoring other objectives or end-to-end network goals. Hence, one of the main challenges that remain at the forefront is that the research on using AI in wireless networks is optimizing one objective while overlooking other constraints such as latency, link, storage, and processing overhead. For example, increasing spectral efficiency using reinforcement learning is proposed in~\cite{6542770}. The cost of information sharing, storing and processing while using the proposed mechanism in real world or large networks is not mentioned. Looking at the overall network performance or end-to-end network objectives, little attention is paid towards a cross-layered approach, as shown in Fig.~\ref{Fig3} in which AI in one layer could also benefit or help in optimization in another layer. Even more so, the negative effects of using AI in one layer over the performance in other layers is rarely considered. For example, the increased latency in finding the optimal route using ML on scheduling in MAC and physical layers are not properly investigated. Due to resource (power, storage, and processing) constraints, a massive number of IoT devices will simply transmit data without performing heavy computation, e.g., for compression or encryption. This will require the upper layers to cooperate to adaptively compress or encrypt data. In mobile IoT nodes, the cross-layer interaction, e.g., for channel or topology selection, from physical to application layer will require synchronization of all layers not only to minimize challenges faced by IoT but also to facilitate end-to-end communication. To properly elaborate the system design complexity, below we describe using ML in digital transceiver design as an example.

\textbf{\textit{The Case of Communication Signal Processing:}}
A communication infrastructure for IoT consists of a central entity, commonly a base station, to handle the traffic of tens or hundreds of IoT devices. The main challenge for the base station is to enable access of a unknown subset of IoT devices at a given instant. Thus, there is a need for efficient signal processing implementation on the base station side. On the other hand, many IoT devices will not require complex signal processing as they do not support multi-antenna communications or complex channel coding. 
Due to the nature of transceiver algorithms, it is difficult to justify the use of ML techniques to replace the conventional signal processing algorithms.  
The current signal processing algorithms are typically designed analytically using statistics, mathematical optimization and information theory. The algorithms based on such techniques are well established and can provide optimal performance. However, many of such algorithms are based on the assumption of a simple and linear system model~\cite{PHYsurvey1}. For example, most beamformer or precoder for a wireless transmitter are based on the assumption that a perfect channel state information (CSI) is available at the transmitter side. However, it is highly unusual to obtain a perfect CSI at the transmitter. In this scenario, a ML based beamformer or precoder can be used which is not dependent on perfect CSI. 
Thus, the application of AI is perfectly justified for transceiver blocks that are highly non-linear in nature and where the mathematical model is far from perfection.
However, there are many sub-optimal solutions available for transceiver algorithms, which are suitable for implementation with satisfactory error-rate performance. The sub-optimal equalization algorithms, such as, zero-forcing, or minimum mean-square error (MMSE) can reach to near-optimal level for massive MIMO systems when the ratio between the number of antennas in a BS and number of users is relatively large~\cite{ADMMIscas}. 
It is difficult to justify the application of AI techniques when a sub-optimal algorithm can provide satisfactory performance. 
To summarize, AI will continue to excel for non-linear signal processing applications like digital pre-distortion which is used to compensate for the non-linearities of a power amplifier. On the other hand, the sub-optimal algorithms can provide very good performance with feasible complexity for many applications. Therefore, more research is necessary to make the AI solutions competitive against those applications.

In most research areas, the processing power required for ML algorithms is not a big hindrance anymore due to the advent of cloud and distributed computing. However, the requirements for digital signal processing are significantly more stringent than traditional applications. Besides, most of the computing required for the physical layer of telecommunications are still carried out by embedded platforms. Some parts of the processing, for example, part of the baseband units, can be transferred to the cloud. In spite of that, the remote radio head (RRU) unit requires highly complex on-site computations, which has to be carried out by embedded computing platforms. 
Thus, the high complexity of sophisticated AI techniques introduces new challenges for the RRUs. 

 We now try to provide an intuitive discussion on the complexity of the neural networks and how they fare against traditional signal processing algorithms with a use case. In general, the Neural Networks require large number matrix calculations. A $N_l=3$ layer fully connected feed-forward NN can be represented as
\begin{equation}
    \mathbf{y} = f(\mathbf{W}_3\,g(\mathbf{W}_2\,h(\mathbf{W}_1\mathbf{x}+\mathbf{b}_1)+\mathbf{b}_2)+\mathbf{b}_3)
\end{equation}
where $h$, $g$ and $f$ are different activation functions for different layers~\cite{NNBook}. The weight matrices for the layers are represented as $\mathbf{W}_1$, $\mathbf{W}_2$ and $\mathbf{W}_3$ and the bias for the layers are represented as $\mathbf{b}_1$, $\mathbf{b}_2$ and $\mathbf{b}_3$. The input and output vectors of this network is denoted as $\mathbf{x}$ and $\mathbf{y}$. It can be seen from the equation that the neural network requires three matrix-vector multiplications. The complexity of a $n\times n$ matrix and $n\times 1$ vectors can be denoted by $n^2$ and thus, the neural network has a $N_ln^2$ complexity if we only consider the matrix multiplications. Here, we assumed each layer has $n$ number of neurons to simplify the comparison.  

The training or learning process to know appropriate weights is a key part of neural networks and the performance of the network is heavily dependent on the methods used for training. The most common training or learning scheme for a NN is known as Backpropagation, which follows Gradient Descent approach that exploits the chain rule. The backpropagation traverses through the same nodes and layers and thus, the number of multiplications after updating the weights is the same as the forward propagation. Therefore, for a 3-layer NN, the total number of operations can be expressed as, $2N_ln^2$. However, the NN requires a large number of iterations of forward and backward propagation to achieve required accuracy for the weights and thus the complexity of the NN training can be expressed as $T(2N_ln^2)$, where $T$ is the number of iterations.   

A least-square solution, which is commonly used in many transceiver operations, requires matrix inversion, which has a complexity of $n^3$ for traditional applications. For large values of $n$, the $n^3$ complexity is higher than the complexity of a forward and backward propagation, i.e. $N_ln^2$~\cite{NNBook}. In spite of the difference, the number of operations for a forward-backward pass of a NN and a matrix inversion is still comparable. 
However, as the value $T$ is typically very large, for example, in hundreds of thousands or in millions, the time required to train a network is too high and impractical. If we take T into account, the numbers of operations are not significantly higher than least-square solutions.
It should be noted that, once trained, the network can run faster than a traditional least-square solution. A NN trained for multi-antenna symbol detection is proposed in~\cite{detnet}. Even though the network can achieve ML performance, it took two days to train the network to properly function as a MIMO detector. These two days required to train the network can render the ML detector useless for many scenarios.

\section{Roadmap: Generalized Global AI Architecture}

\begin{figure*}[h]
\centering
\includegraphics[keepaspectratio,width=1.80\columnwidth]{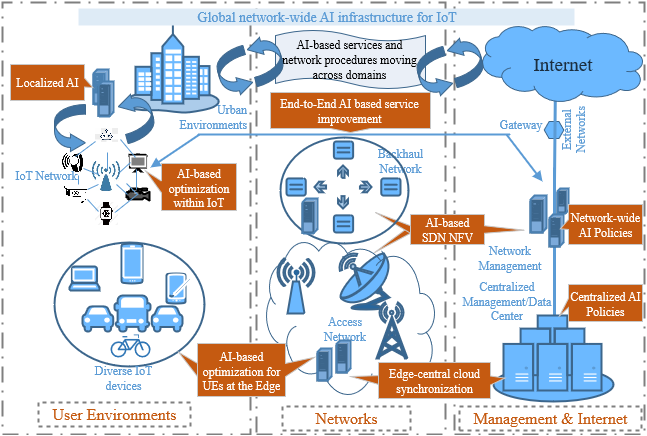}
\caption{AI operations in an AI-based communication network infrastructure.}
\label{NetArch}
\end{figure*}

Even though there exists a plethora of research on using AI in communication networks for different use-cases, applications, network functions and segments, little efforts are put on visualizing the holistic network architecture. The major benefit of the holistic network view is to attain the end-to-end goals without having situations where achieving one objective leads to a compromise on another. In addition, having a global network view is vital to efficient utilization of available resources throughout a network. Therefore, a global network architecture using AI is presented in Fig.~\ref{NetArch}. The three-tier architecture represents autonomous and intelligent network operations leveraging AI in each tier, as well as across the three tiers in order to maintain synchronized AI-based operations in the entire network for different IoT services.

\begin{figure}[h]
\centering
\includegraphics[keepaspectratio,width=1.0\columnwidth]{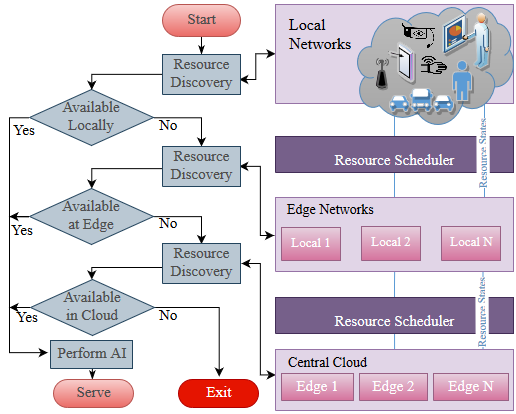}
\caption{Resource discovery for AI based operations.}
\label{Resource}
\end{figure}

In Fig.~\ref{NetArch}, the user environments comprises end-user devices (IoT devices) and IoT networks that use AI to improve its performance. Due to limitations of IoT devices such as processing and storage, edge (or MEC) platforms are used when higher resources for AI operations are needed. Since edge platforms still represent distributed operations with limited capabilities, centralized cloud systems are proposed for two major reasons. First, to maintain global network view including AI operations in order to maintain synchronized operations throughout the network. Second, to provide higher resources when the edge platforms fall short of resources. On one hand, the communication network infrastructure using diverse technologies from radio access technologies to the application layer facilitates AI operations throughout the network, i.e., from end-user environments to the centralized cloud systems. On the other hand, AI is used in the communication network infrastructure to improve end-to-end goals of the network. Therefore, it can be seen in Fig.~\ref{NetArch} that AI is used through out the network, connecting many IoT networks, edge platforms and centralized cloud systems. Fig.~\ref{Resource} represents the three tier network architecture visualizing how in practice AI will be used in a large network, such as shown in Fig.~\ref{NetArch}. 

In Fig.~\ref{Resource}, the local network represents a network of IoT devices, which in most cases have very limited resources, e.g., storage, computing and transceiver capabilities. Edge networks comprise edge nodes, each node is near to an IoT network to meet latency constraints, and have more resources compared to local IoT networks. The centralized cloud serves many edge networks and have higher resources to serve the entire (global) network. The resources available within each network, i.e. local IoT and edge networks, are visible in the centralized control system, much like the data plane resources being visible in the SDN control plane. Since the tools and algorithms of AI require data, and a common assumption is that the more the data is, the better the results will be~\cite{7906512}, it is highly likely that the needed resources are not available locally or in the edge. In that case, requests must be sent to high-resourced centralized cloud systems to fulfill the requirements of processing and storage. Thus, AI procedures throughout the entire network are carried out in the following three steps: 

\begin{enumerate}
  \item Local resource discovery: Before initializing an AI procedure, a local resource discovery procedure is carried out. If the resources, such as storage and computing, are available locally, the process is carried out within the local IoT network. 
  \item Edge resource discovery: If the local IoT network resources are not enough, the resource discovery procedure in the edge layer will be carried out. If the resources in the edge are enough and are available for the AI procedure, edge resources will be allocated and all the processing will happen in the edge layer. It is important to note that link and bandwidth resources will also count for sending the (possibly raw) data, training parameters, and decisions back and forth between the local IoT network and the edge nodes. 
  \item Central cloud resource discovery: If the edge resources fall short, the resource discovery and allocation procedure will be carried out in the central cloud. Thus, the AI processing will happen in the centralized cloud, and even more network resource will be consumed in this case. 
\end{enumerate}

Such globally optimized network architecture will have the potential to avoid many challenges described in the previous section. For example, localized IoT-based AI processing yields benefits such as low bandwidth consumption and meeting the requirements of latency, as evaluated for wireless sensor networks in~\cite{Branch2013}. The edge nodes are involved for two major reasons, first, the resources in IoT or local networks are not capable (fall short) to perform the tasks, and second, the latency constraints do not allow to perform the tasks in the centralized cloud systems as elaborated in~\cite{8270639}. Yet, the edge nodes have challenges in terms of resources mainly due to the humongous growth of data as well as distributed ML required for distributed services that require global aggregation of, for instance, data and learning parameters~\cite{8486403}. Since most of the IoT environments are dynamic and evolving, the learning models are bound to evolve, which can also create challenges in synchronizing and monitoring multiple edge nodes~\cite{Soto:2016:CMM:2991561.2991575}. Therefore, a dynamic architecture capable to synchronize multiple edge nodes through a centralized control and monitoring system, as depicted in Fig.~\ref{NetArch}, is required. In the proposed architecture, the concepts of intelligent service decoupling for enabling mobility of AI systems (e.g. AI system running as a virtual function) between multiple edge nodes, and between edge nodes and centralized monitoring and control systems is visualized. Intelligent AI services for IoT can be decoupled much like Network Function Virtualization (NFV) with the unique requirements of IoT and AI, as elaborated in~\cite{ramos2019intelligence}. AI system or service mobility, along with synchronizing the needed AI processing among multiple edge nodes~\cite{8486403}, can be achieved through the novel technological development in communication and computing technologies such as SDN and MEC, as explained with examples in~\cite{7954011}. The hierarchical architecture is beneficial in terms of proactive catching without draining the local and edge resources or compromising the latency constraints, as evaluated in~\cite{8299576}.

\section{Discussion and Future Research Directions}

AI with its many disciplines, tools and algorithms, will play an important role to efficiently utilize the available network resources for IoT through autonomous network operations. However, deploying AI mechanisms need proper investigation of the resulting consequences in terms of different performance indicators. Furthermore, the effects of using AI in one service over the other, and one network segment for function over the other must also be properly investigated. For example, raw data gathering, processing it, and disseminating the resulting information or decisions of AI can increase communication overhead resulting in network congestion, or induce delays in different network functions such as routing or access control.

Therefore, it is highly important to investigate the resulting challenges in the underlying network infrastructure due to integrating the mechanisms of AI in communication network infrastructures that will be used by IoT. The major challenges discussed throughout this article are summarized in Table II with the most important references. The challenges are presented with respect to using AI within IoT devices, within a localized network of IoT, edge level AI that runs AI procedures in the edge nodes, and centralized cloud level AI in which high performance cloud infrastructures are used for AI processing. The challenges are given different levels, from low, medium and high, to give an insight into its severity, based on the references.

\begin{singlespace}


\makeatletter
\newcommand{\thickhline}{%
    \noalign {\ifnum 0=`}\fi \hrule height 1pt
    \futurelet \reserved@a \@xhline
}

\begin{table*}[h]
\caption{Summary of challenges in wireless network infrastructure leveraging AI for IoT.}
\centering
\begin{tabular}{|p{3.8 cm} |p{1.80 cm}  |p{2.0 cm} |p{2.2 cm} |p{2.2 cm}  | p{3.5 cm}  |} 
\hline

\multirow{2}{*}{\textbf{Challenges}} & \multicolumn{4}{|c|}{\textbf{IoT Environment}} & \multirow{2}{*}{\textbf{References}}\ \\ [0.5ex] 
\cline{2-5}
         & IoT Device  & Localized AI & Edge Level AI & Cloud Level AI &  \\ [0.5ex] 
\thickhline

Communication Overhead & Low &Low& Medium& High &~\cite{8489985},~\cite{8664630},~\cite{Arjevani2015},~\cite{7906512}\ \\ \hline
End-to-End Latency & Low & Low & Medium & High & ~\cite{8057306},~\cite{8227766},~\cite{8279592},~\cite{8647927},~\cite{OffloadIoT}\ \\ \hline
Security Challenges& Low& High& High &High & ~\cite{Huang2011},~\cite{5504793},~\cite{Martin}\ \\ \hline
Caching and Memory & High & High & Medium & Low &~\cite{8192463},~\cite{Suthaharan} \ \\ \hline
Network Traffic Control & High & High & Medium & Low & ~\cite{7792369},~\cite{7935536},~\cite{Suchara:2011:NAJ:1993744.1993756}\ \\ \hline
System Complexity & High & High & Medium & Low & ~\cite{detnet},~\cite{PHYsurvey1}\ \\ \hline


\thickhline
\hline
\end{tabular}
\renewcommand{\arraystretch}{2}

\begin{tablenotes}
            \item *Challenge levels or severity represented by Low, Medium, and High.
\end{tablenotes}

\end{table*}

\end{singlespace}

Certain measures taken according to the context, the network infrastructure, and available resources can help us use the mechanisms of AI more effectively. Comparing the requirements, for instance, time-sensitivity of applications vs benefits of using the disciplines of AI either in the local IoT network or in the centralized cloud systems might provide better conclusions. For example, latency-critical applications need to use the concepts of service migration (AI processing) from the central cloud to edge or local IoT gateways. In this case, AI must be bundled with efficient service migration techniques and slice elasticity to increase or decrease resources (e.g. in the edge nodes) accordingly. Therefore, the holistic view of the global network infrastructure and available resources will be highly beneficial. However, further research is necessary in the following communication network-specific areas to reap the full benefits of AI.

\textbf{\textit{AI-based approaches within the bandwidth, spectrum, and latency constraints:}} It is foreseeable that the number of end-user devices will grow exponentially in future wireless networks, having different traffic patterns, and mostly prone to security challenges~\cite{8086136}. Hence, how to efficiently use the existing allocated bandwidth and spectrum resources while not compromising on the required data rates, QoS and QoE, will be a huge challenge. AI based approaches, which can predict the traffic growth and flash traffic, proactively move services between edge and centralized cloud systems, and dynamically allocate resources will definitely yield better results. However, more research is needed to develop AI mechanisms that can be trained quickly and effectively with less data in order to consume less bandwidth or spectrum resources.

\textbf{\textit{AI-based security approaches for AI-based security challenges:}} Conventionally, using AI for improving network security is highly researched, but on the contrary, e.g., using AI for security attacks on network entities, must also be investigated. Security attacks leveraging AI can be more challenging to detect, or stop as demonstrated in~\cite{Martin}, and described in~\cite{Huang2011}. Similarly, AI needs data, and data needs privacy. Therefore, AI based approaches to secure resources and data from AI based security threats and privacy issues represent interesting challenges that need further research.

\textbf{\textit{Meeting the caching requirements in times of Big Data:}} Using the tools of Big Data analytics require resources such as storage, computing and link capacities. However, finding early enough if the data can be counted as Big Data will lead better selection of technologies for the purpose, as described in~\cite{Suthaharan}. Now that resources near the users, i.e., edge clouds, are gaining footprints in communication networks, scaling the resources up for Big Data will eventually not pose major challenges. However, further research is required to enable using AI on Big Data near near the data sources within the resource constraints.

\textbf{\textit{Network abstraction to cope with complexity:}} Abstracting the underlying network infrastructure, from services that will use it, will simplify the network to be used for any kind of services. Granular, event-driven control of the network elements through high-level policies, and avoiding low-level configurations has been enabled by SDN~\cite{6739370}. The same mechanisms, i.e., functional split and abstraction, have been proposed for the radio access technologies, however, further research is required on the MIMO side. Abstraction in IoT has driven research and industry interests also, as seen from the Pelion IoT platform. Leveraging AI in the same fashion, AI-as-a-service whenever and wherever needed in a network, will result in the same benefits without increasing complexity of the overall system, which needs further research.

\section{Conclusion}

AI has gained a research momentum in wireless networks to cope with the increasingly complex nature of diverse IoT devices and services. However, most state-of-the-art research takes the concepts of AI from other mature technologies such as robotics and computer vision \textit{as it is} and use it to solve different complex challenges faced by IoT devices and services, as well as the underlying network serving IoT. Such right-away use of the concepts of AI in the wireless network infrastructure gives rise to many challenges. In this article, the main challenges are highlighted with potential solutions and open research issues that need further research. The main objective of this work is to drive attention for future research towards wireless network-specific design of the concepts, tools, algorithms, and even disciplines of AI for the communication of IoT. Furthermore, generic requirements of an IoT wireless network are highlighted to elaborate the need and integration points of the concepts of AI into the wireless network infrastructure used by IoT. The challenges arising in each integration point of AI and wireless networks are discussed. A generalized conceptual framework, as a roadmap, is suggested that could solve most of the challenges with novel technological concepts used for network programmability, global network resource visibility, and granular control of network and AI functions.



\ifCLASSOPTIONcaptionsoff
  \newpage
\fi

\begin{singlespace}

\bibliographystyle{IEEEtran}
\bibliography{IEEEbibl}

\end{singlespace}

\end{document}